\documentclass[aps,prl,twocolumn,groupedaddress,showpacs]{revtex4}

\usepackage{epsfig}
\begin{document}

\title{Magnonic Analogue of Black/White Hole Horizon in Superfluid $^3$He-B}

\author{M. \v{C}love\v{c}ko}
\author{E. Ga\v{z}o}
\author{M. Kupka}
\author{P. Skyba}
\email[Electronic Address: ]{skyba@saske.sk}

\affiliation{{Institute  of Experimental Physics, SAS} and P. J. \v{S}af\'arik University
Ko\v{s}ice, Watsonova 47, 04001 Ko\v{s}ice, Slovakia.}

\date{\today}

\begin{abstract}
We report on theoretical model and experimental results of the  experiment made in a
limit of absolute zero temperature ($\sim$ 600\,$\mu$K) studying the spin wave analogue
of black/white hole horizon using  spin (magnonic) superfluidity in superfluid $^3$He-B.
As an experimental tool simulating the properties of the black/white horizon we used the
spin-precession waves propagating on the background of the spin super-currents between
two Bose-Einstein condensates of magnons in form of homogeneously precessing domains. We
provide experimental evidence of the white hole formation for spin precession waves in
this system, together with observation of an amplification effect. Moreover, the
estimated temperature of the spontaneous Hawking radiation in this system is about four
orders of magnitude lower than the system's background temperature what makes it a
promising tool to study the effect of spontaneous Hawking radiation.

\end{abstract}


\maketitle

Recently, it has been shown that Hawking radiation should be related not only to physics
of the astronomical black holes \cite{hawking}, but should be viewed as a general
phenomenon of a dynamic nature of various physical systems having capability, at certain
conditions, to create  and form a boundary - an event horizon \cite{tutorial}. According
to theory, a fundamental dynamical property of any event horizon analogue is a
spontaneous emission of thermal Hawking radiation, the temperature of which depends on a
velocity gradient at the horizon \cite{unruh1,ralf1}
\begin{equation}
T=\frac{\hbar}{2 \pi k_B}\frac{\partial v^{r}}{\partial r} \sim 10^{-12} \frac{\partial v^{r}}{\partial r}.
\end{equation}
Belong a set of physical systems used to model black/white hole horizon can be included:
the sound waves in trans-sonic fluid flow \cite{unruh1,visser},  the surface waves on
flowing fluid \cite{ralf,rousseaux1,rousseaux2,silke},  hydraulic jumps in flowing
liquids \cite{volovik, jannes}, and light in optical fibre \cite{philbin}. It turns out,
however, that temperature of the spontaneous Hawking radiation is typically a several
orders of magnitude lower than the background temperature of the physical systems used as
an experimental tool to study this phenomenon. Perhaps, solution to this problem is to
find another, suitable condensed matter systems, which can model the event horizon, but
with background temperature approaching absolute zero temperature
\cite{BEC1,BEC2,BEC3,magnonic}.

\begin{figure}[t!]
\begin{center}
\epsfig{figure=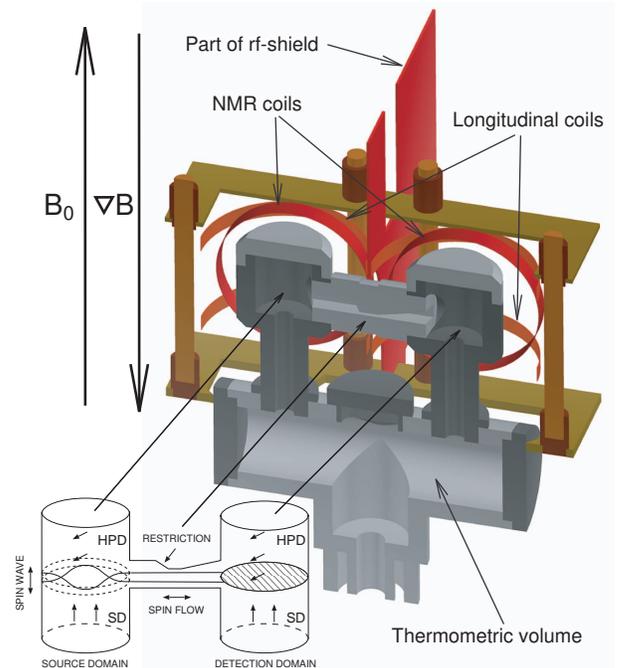,width=80mm} \caption{(Color on line) Schematic 3-D cross section of
the experimental cell and concept of the experiment. The channel dimensions are: the width is 3\,mm,
the height is  0.4\,mm and the channel length is 2\,mm.} \label{fig1}
\end{center}
\end{figure}
In this Letter we present as a theoretical model, so experimental results of the first
experiment made in a limit of absolute zero temperature ($\sim$ 600$\mu$K) studying  the
black/white hole horizon analogue using a physical system based on the spin (magnonic)
superfluidity in superfluid $^3$He-B. The concept of the experiment is quite simple
\cite{spinhole}. The experimental cell with superfluid $^3$He-B consists of two cylinders
mutually connected by a channel (see Fig.~\ref{fig1}). The cell is placed in steady
magnetic field $B_0$ and magnetic field gradient $\nabla$B, both oriented along $z$-axis.
Using cw-NMR technique, we created a Bose-Einstein condensate of magnons in a form of the
homogeneously precessing domain (HPD) in both cylinders \cite{hpd1,fomin,jltp1997}. The
HPD is a dynamical spins structure formed in a part of the cell placed in lower magnetic
field which, with an aid of the dipole-dipole interaction, coherently precess around a
steady magnetic field at the angular frequency $\omega_{rf}$. Within the rest of cell,
the spins are co-directional with the steady magnetic field and do not precess. These two
spin domains are separated by a planar domain wall, the position of which is determined
by the Larmor resonance condition $\omega_{rf}= \gamma (B_0 + \nabla B.z)$, where
$\gamma$ is the gyromagnetic ratio of the $^3$He nuclei.

To model black/white hole horizon in superfluid $^3$He-B, we used two fundamental
physical properties of the HPD: the spin superfluidity  and the presence of the HPD's
collective oscillation modes in a form of the spin precession waves
\cite{euro1997,prl2003,prl2008,prb2012}. The spin superfluidity allows to create and
manipulate the spin flow, i.e. the spin super-currents flowing between precessing domains
as a consequence of the phase difference $\Delta \alpha_{rf}$ between the phases of spin
precession in individual domains \cite{moscow}. The spin precession waves serve as a
probe testing formation and presence of a black/white hole horizon inside the channel:
channel has a restriction allowing to reach the different regimes of the velocity of the
spin super-current flow with respect to the group velocity of the travelling
spin-precession waves. This is similar to the experiment suggested by R. Sch\"{u}tzhold
and W. Unruh \cite{ralf} and later performed by G. Rousseaux et al.
\cite{rousseaux1,rousseaux2,silke} and S. Weinfurtner et al. \cite{silke}. In the
presented experiment, the magnetic field gradient $\nabla$B plays the role of the
gravitational acceleration, the spin super-currents represent the water flow, and the
spin-precession waves correspond to the gravity waves on the water's surface.


In order to develop a mathematical model, we initially consider a simplified problem - a
volume of $^3$He-B placed into a large steady magnetic field $B_0$ with gradient field
$\nabla B$ applied in the $z$-direction and a small magnetic rf-field $B_{rf}$, which
rotates in the ``horizontal" $x-y$ plane at the angular frequency $\omega_{rf}$, with the
phase of rotation varying linearly with $x$. Cartesian components of the resultant
magnetic field are $B_x= - B_{rf}\cos (\omega_{rf}t +\nabla \alpha_{rf}x)$,
$B_y=B_{rf}\sin (\omega_{rf}t + \nabla \alpha_{rf}x)$ and $B_z=-B_0 + \nabla B.z$. This
model problem is treated theoretically by adapting the method we presented elsewhere
\cite{prb2012}.

The steady-state response is a layer of magnetization  precessing with the frequency and
local phase of the rf-field lying over a layer of stationary magnetization. The domain
with precessing spins (magnetization) may oscillate about its steady state. The principal
variable describing small oscillations is the perturbation   $a(t,x,y,z)$ to the phase of
spin precession. In the long wavelength approximation and for a thin domain of precessing
spins, the perturbation propagating along the domain wall is found to be governed by the
equation

\begin{center}
\begin{eqnarray}
\left(\frac{\partial }{\partial t} - \frac{\partial }{\partial x}\,u\right)
\left(\frac{\partial }{\partial t} - u\,\frac{\partial }{\partial x}\right) a_{dw}
-\frac{\partial}{\partial x}\left( c^2 \frac{\partial a_{dw}}{\partial x}\right)\nonumber \\
-\frac{\partial}{\partial y}\left( c^2 \frac{\partial a_{dw}}{\partial y}\right)
+ \sqrt{\frac{3}{5}}\, \gamma^2 \nabla B B_{rf}L a_{dw}=0.
\label{equ1}
\end{eqnarray}
\end{center}
Here, $a_{dw}(t,x,y) = a(t,x,y, z_{dw})$, where  $z_{dw}$ denotes $z$-coordinate of the
domain wall position and  $L$ is the thickness of precessing domain.  Two terms $u$ and
$c$ represent the spin flow and the group wave velocities, respectively, and they can be
expressed as
\begin{equation}\label{equ2}
u=\frac{(5c_L^2-c_T^2)\,\nabla \alpha_{rf}}{2 \omega_{rf}}, \hspace{3mm} c^2=\frac{(5c_L^2+3c_T^2)\,\gamma \nabla B L}{4\omega_{rf}}.
\end{equation}
where and $c_L$ and $c_T$ denote the longitudinal and transverse spin wave velocities
with respect to the field orientation, respectively. We shall assume that $\nabla
\alpha_{rf} $ is localized on the length of the sharpest restriction in the channel of
the order $dl$ = 0.5\,mm, therefore $\nabla \alpha_{rf} \sim \Delta \alpha_{rf} /dl$.

The long spin-precession waves travelling along the surface of a thin layer of precessing
and flowing spins are governed by the same equation as a scalar field in a
(2+1)-dimensional curved space-time.  Thus, these waves experience the background as an
effective space-time with the effective metric
\begin{equation}\label{equ3}
ds^2 \,=\,  c^2\left[-c^2dt^2+ (dx+udt)^2+dy^2\right].
\end{equation}
As it is implied by this equation, an ``event horizon" for the long spin-precession waves
is formed where and when $u^2=c^2$. For the sake of completeness, the exact dependence of
the angular frequency $\omega$ of any spin-precession wave on components $k_x$ and $k_y$
of its wave vector is
\begin{equation} \label{equ4}
\left( \omega + u\,k_x\right)^2=\frac{\gamma \nabla B}{2 \omega_{rf}}c_1^2 \kappa \tanh(\kappa L),
\end{equation}
where $c_1^2 = (5c_T^2-c_L^2)$ and $\kappa$ is defined as
\begin{equation}\label{equ5}
\kappa^2 = \frac{3}{2c_1^2}\left[ \frac{4}{\sqrt{15}}\omega_{rf}\gamma \,B_{rf} +\frac{1}{3}(5c_L^2+3c_T^2)(k_x^2+k_y^2)\right].
\end{equation}
It is important to note that the applied rf-field explicitly determines the ``vacuum",
that is to say, it prescribes the angular frequency and the variation of the phase for
the precessing magnetization (spins) representing a steady state of the system
considered. But Eq.~(\ref{equ1}) for small oscillations superposed on the steady state is
determined primarily by the properties of that state, no matter how the steady state was
created (except for the ``mass" term that is determined by the external field
explicitly). So, the Eq.~(\ref{equ1}) is capable of describing the perturbations of the
steady state in the absence of rf-field, if the precession of background magnetization
has maintained its given angular frequency and phase. Although derived for a background
represented by a uniform magnetization flow parallel to the flat top of the cell, the
above presented relations can be used for qualitative analysis and quantitative
estimation of the situation where $\nabla \alpha_{rf}$ and $L$ slightly vary on spatial
scales larger than $L$.


As mentioned above, we performed the experiment in the cell shown in Fig. \ref{fig1}. The
cell was attached to Ko\v{s}ice's diffusion nuclear stage \cite{stage}, filled with
$^3$He at 3\,bars and using the adiabatic demagnetization cooled down to temperature of
$\sim$ 0.5T$_c$. The temperature of the $^3$He was measured using a powder  Pt-NMR
thermometer immersed in the liquid and calibrated against the $^3$He superfluid
transition temperature T$_c$. The HPDs were simultaneously and independently excited in
both cells using cw-NMR method at angular frequency
$\omega_{rf}=2\pi.462\times10^3$\,rad/sec. To achieve this, we used two rf-generators
working in phase-locked  mode  and with zero phase difference $\Delta\alpha_{rf}$ between
excitation signals. The induced voltage signals from NMR coils were amplified by
preamplifiers and measured by two rf-lock-in amplifiers, each controlled by its own
generator. In order to reduce the mutual crosstalk between the rf-coils, each rf-coil was
covered by a shield made of a copper foil. The longitudinal coils provided an additional
alternating magnetic field used for the spin-precession waves generation \cite{prl2003}.

\begin{figure}[t!]
\begin{center}
\epsfig{figure=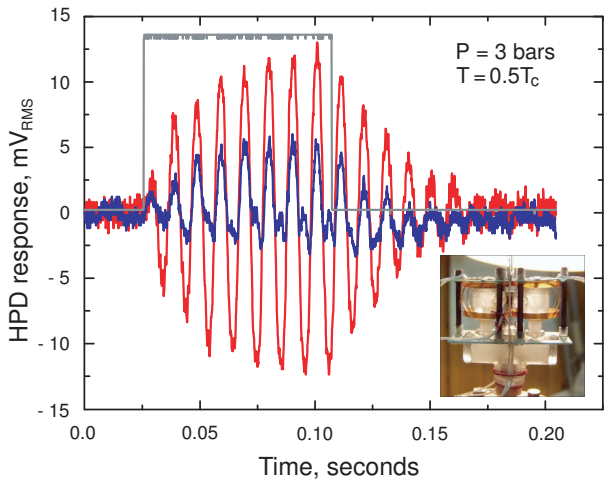,width=80mm} \caption{(Color on line) Voltage signals
corresponding to spin-precession waves: the source domain (red), the detection domain (blue).
The rectangular signal shows the time window when 8 sinusoidal excitation pulses were applied in
order to excite the spin-precession waves. Insets: the picture of the experimental cell on the
bench before rf-shield installation.} \label{fig2}
\end{center}
\end{figure}

Once two HPDs were generated, the position of the domain wall was adjusted into channel.
This step is easy to accomplish by means of the homogeneous field $B_0$ (with aid of a
small longitudinal oscillations), as the position of the domain wall follows the plane
where the Larmor resonance condition is satisfied. Specifically, the precision of the
domain wall  adjustment is given by $\Delta B_0/ \nabla B $, where $\Delta B_0 =
0.76\,\mu$T is the field step controlled by the current source and for $\nabla B$=15 mT/m
giving a spatial precision of $\sim$ 50 $\mu$m \cite{source}. For comparison, the domain
wall thickness $(\lambda_F = c_L^{2/3}/(\gamma \nabla B \omega_{rf})^{1/3})$ for the
above parameters is $\sim$ 0.34\,mm. As the height of the channel is 0.4\,mm, an
estimated length of the precessing layer $L$ in the channel is $L\sim$ 150\,$\mu$m $\pm$
50 $\mu$m. The spin flow between HPDs can be established in both directions depending on
the sign of the phase difference $\Delta\alpha_{rf}$, while the spin flow velocity $u$
depended on the magnitude of $\Delta\alpha_{rf}$. The details of the experiment are
provided in \cite{prb2019}.

The spin-precession waves in the source domain were generated by 8 sinusoidal pulses at
an appropriate low frequency using the separate generator. The low frequency response
from the source and detection HPD for a particular value of $\Delta \alpha_{rf}$ was
extracted from rf-signal by a technique based on application of a rf-detector and a
low-frequency filter. The low frequency signals were stored by a digital oscilloscope for
the data analysis. The examples of the signals representing the excited spin-precession
wave in the source domain and incoming wave in the detection  domain are showed in
Fig.~\ref{fig2}. When the pulse is finished, there are clear free decay signals of the
spin-precession waves from both domains, and these parts of the signals were analyzed by
the methods of the spectral analysis as a function of the phase difference $\Delta
\alpha_{rf}$, i.e. as a function of the spin flow velocity $u$.

\begin{figure}[htb]
\begin{center}
\epsfig{figure=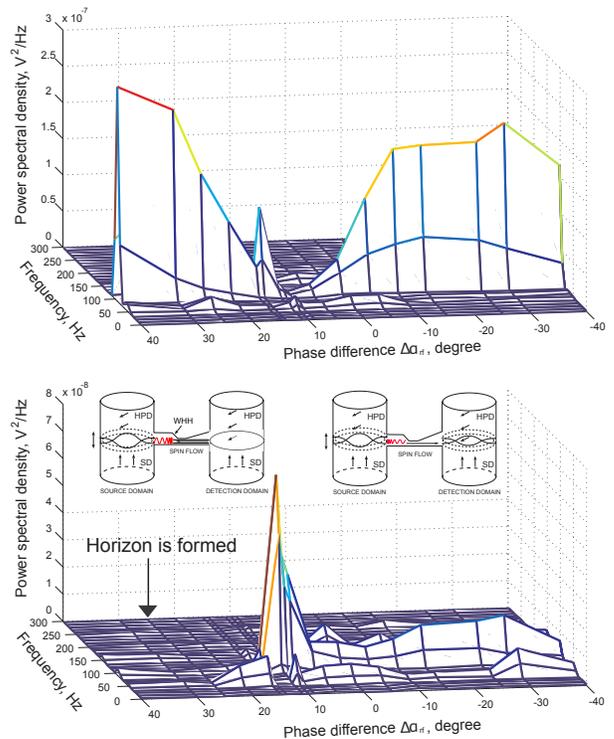,width=90mm} \caption{(Color on line) The power spectral
density (PSD) of the free decay signals as function of the phase difference measured from the source domain
(upper) and the detection domain (bottom). Insets show a schematic illustration
of the spin waves dynamics in channel.} \label{fig3}
\end{center}
\end{figure}

Figure \ref{fig3}  shows the power spectral density (PSD) of the free decay signals for
the source (upper) and detection (lower) domains as a function of the phase difference
$\Delta \alpha_{rf}$. There are a few remarkable features presented there. Firstly, there
are relatively strong PSD signals in the source domain with an exception of a deep
minimum at region of $\Delta \alpha_{rf}$ corresponding to $\sim$ 10$^\circ$. Secondly,
the weaker PSD signals are  observed in the detection domain in the range of negative
values of $\Delta \alpha_{rf}$ up to 10$^\circ$, above which strong PSD signals were
measured. Thirdly, no PSD signals within experimental resolution were measured in the
detection domain for values of $\Delta \alpha_{rf} \gtrsim$ 25$^\circ$. How can one
interpret these data?

For the negative values of $\Delta \alpha_{rf}$, the spin super-currents flow from the
source domain towards the detection one. Thus, the spin-precession waves excited in the
source domain are dragged by the spin super-currents and they travel downstream to the
detection domain, where they are detected. The amplitude of the detected waves is reduced
by process of the energy dissipation inside channel due to spin diffusion, and by the
spin flow modifying the frequency of spin-precession waves which is slightly different
from the resonance frequency of the standing waves.

The deep signal minimum in the source domain and corresponding maximum in the detection
domain for $\Delta \alpha_{rf} \sim$ 10$^\circ$ - 15$^\circ$ is consequence of zero spin
flow between domains that  leads to a resonance match \cite{ccc}. Therefore, when the
spin-precession waves are excited in the source domain at this condition, all energy is
transferred to and absorbed by the detection domain at once. For  $\Delta \alpha_{rf}\,>$
10$^\circ$ the direction of the spin flow is reversed, i.e. the spin super-currents flow
towards the source domain and the emitted spin-precession waves propagate against this
flow. The change in direction of the spin super-currents is also seen on the phase of the
decay signal from the source domain as the gradual phase shift by ~180$^\circ$
\cite{prb2018}.

As one can see from Fig. \ref{fig3}, there are no PSD signals detected in the detection
domain for $\Delta \alpha_{rf} \gtrsim$ 25$^\circ$. We interpret this as a formation of
the white hole horizon in the channel: spin-precession waves sent from the source domain
towards to the detection domain are blocked by the spin flow and never reach the
detection domain. This interpretation is supported by calculation using above presented
model: the white hole horizon is formed in a place, where and when the  condition
$c^2=u^2$ is satisfied. For given experimental parameters and assuming that $\nabla
\alpha_{rf}$  is localized on the length of $dl$ = 0.5\,mm, in order to satisfy the
condition $c^2=u^2$ for the phase difference $\Delta \alpha_{rf} \sim 30^{\circ}$,
estimated length of the precessing layer $L$ is $L\sim$ 100\,$\mu$m, what reasonably
corresponds to the experimental value \cite{prb2018}.

\begin{figure}[ht!]
\begin{center}
\epsfig{figure=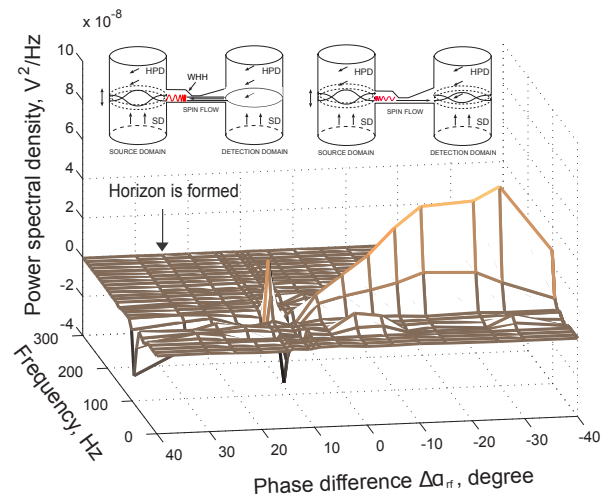,width=85mm} \caption{(Colour on line) The cross-correlation  power spectral
density of the source and detector free decay signals as function of
the phase difference $\Delta \alpha_{rf}$.} \label{fig4}
\end{center}
\end{figure}

Finally, Fig.~\ref{fig4} shows the cross power spectral density between free decay
signals measured in source and detector domains as function of the phase difference
$\Delta \alpha_{rf}$. For values of $\Delta \alpha_{rf} <$ 0$^\circ$, i.e. when the spin
flow drags excited spin-precession waves from the source domain towards to the detection
domain, the decay signals from both domains are correlated. For values of $\Delta
\alpha_{rf}\,>$ 0$^\circ$, reduction and and following reversion of the spin flow affects
the dynamics of the propagation of the spin-precession waves that leads to the change in
correlation between the decay signals detected in both domains. When the spin flow
approaches zero value, due to resonance match between the domains, the energy is
transferred in both directions that is manifested as correlation/anti-correlation peaks.
However, when the white horizon is formed ($\Delta \alpha_{rf} >$25$^\circ$), the decay
signals are anti-correlated. We may interpret this in way that the rise of the decay
signal in the source domain is paid by the spin flow flowing from the detection domain
towards to the source domain - in agreement with  theoretically predicted amplification
of the wave on the horizon paid by the energy of the flow \cite{unruh2}. This
interpretation is supported by dependence presented in Fig. \ref{fig3} (upper
dependence), where a notable feature  regarding to the absolute values is showed: when
spin current flows from the detection domain to the source domain, the waves in source
domain  have a tendency to have a higher power spectral density amplitude than those for
the spin current flowing in opposite direction. However, to confirm the physical origin
of the observed phenomena additional measurement have to be done.

In conclusion, we performed the experiment in a limit of absolute zero temperature
probing the black/white hole horizon analogues in superfluid $^3$He-B using the
spin-precession waves propagating on the background of the spin super-currents between
two mutually connected HPDs and provided the evidence of the white hole formation for
spin precession waves. Moreover, the presented theoretical model and experimental results
demonstrate that the spin-precession waves propagating on the background of the spin flow
between two HPDs possess all physical features needed to elucidate physics associated
with the presence of the event horizons, e.g. to test the spontaneous Hawking process. In
fact, assuming that the spin super-currents velocity of the order of $u \sim $ 1\,m/s
varies on the length of $dl\sim 10^{-4}$\,m one can estimate the temperature of the
Hawking radiation in this system to be of the order of 10 nK, what is a temperature only
four orders of magnitude lower that the background temperature, and this makes presented
system a promising tool to investigate this radiation \cite{prb2018}.

We acknowledge support from European Microkelvin Platform (H2020 project 824109),
APVV-14-0605, VEGA-0128 and ERDF-ITMS 26220120047 (Extrem-II). We wish to thank \v{S}.
Bic\'ak and G. Prist\'a\v{s} for technical support. Support provided by the U.S. Steel
Ko\v{s}ice s.r.o. is also gratefully acknowledged.


\begin{thebibliography}{99}
\bibitem{ccc} The 10$^\circ$  shift from $\Delta \alpha_{rf}$ = 0 for zero spin flow is
    caused by small discrepancy in the resonance frequencies settings of two resonance
    circuits used for HPD excitation.
\end{thebibliography}
\end{document}